\documentclass[11pt,a4paper]{article}
\pdfoutput=1
\usepackage{jheppub}
\usepackage{subfig}

\newcommand{\ba}{\begin{eqnarray}}
\newcommand{\ea}{\end{eqnarray}}

\newcommand{\barray}{\begin{array}}
\newcommand{\earray}{\end{array}}

\def\del{\partial}

\title{Recrudescence of massive fermion production by oscillons}

\author{Paul M. Saffin,}

\affiliation{School of Physics and Astronomy, University Park, University of Nottingham,\\ Nottingham NG7 2RD, United Kingdom}

\emailAdd{paul.saffin@nottingham.ac.uk}

\abstract{We bring together the physics of preheating, following a period of inflation, and the dynamics of non-topological solitons, namely oscillons. We show that the oscillating condensate that makes up an oscillon can be an efficient engine for producing heavy fermions, just as a homogeneous condensate is known for doing the same. This then allows heavy fermions to be produced when the energy scale of the Universe has dropped below the scale naturally associated to the fermions. }


\begin{document}

\maketitle

\section{Introduction}
\label{sec:intro}

The period in the early Universe following an inflationary epoch can be a rather explosive time. In many models of inflation the Universe transforms from a cold, empty, near de Sitter phase into a hot environment, via some sort of resonant behaviour of the fields \cite{Dolgov:1989us}.
The basic idea is that the inflaton forms a (nearly) homogeneous oscillating condensate and some modes of a daughter species, to which the inflaton is coupled, are in resonance with the oscillations, causing their particle number to grow exponentially. 
This (p)reheating scenario has found many uses in early-Universe cosmology, including the production of gravitational waves \cite{Khlebnikov:1997di};
non-thermal phase transitions and the formation of topological  \cite{Kasuya:1997ha}
and non-topological defects \cite{Amin:2010xe}
\cite{Amin:2011hj}; 
baryogenesis \cite{Kolb:1996jt}.



This great enhancement in particle number is not available to fermions, as they must satisfy the Pauli exclusion principle, which limits their number occupancy to unity. This Pauli-blocking was initially believed to mean that a perturbative calculation of fermion effects would suffice, and that their presence would not significantly affect the inflaton decay \cite{Dolgov:1989us}. The physics of such a decay process ignores the coherent nature of the inflaton and treats the fermion production as coming from the decay of single inflaton particles, leading to a decay rate of $\Gamma_{\phi\to\psi\psi}=\frac{\xi^2m_\phi}{8\pi}$, and a narrow peak in the spectrum centred around $m_\phi/2$, where $\xi$ is the Yukawa coupling and $m_\phi$ the inflaton mass \cite{Dolgov:1989us}. However, the equation of motion that governs the mode functions for the fermion field are, in fact, rather similar to those that appear in bosonic preheating, and while they do indeed exclude the possibility of the particle number exceeding unity, it is clear that the perturbative approach is insufficient, as noted by \cite{Boyanovsky:1995ema}. The importance of this departure from the perturbative result was stressed in \cite{Baacke:1998di}, where it was found that fermions could get excited within tens of inflaton oscillations, as opposed to the $\sim10^{14}$ predicted by the perturbative calculation.

A further distinction between bosonic and fermionic preheating is that the effective mass of fermions can vanish as the condensate evolves through certain values, making their creation much easier than bosons (see \cite{Tkachev:2001zi} for a clear comparison of fermion and boson production due to preheating). This effect was utilised for leptogenesis in \cite{Giudice:1999fb} to produce massive right-handed neutrinos, orders of magnitude heavier than the oscillating inflaton field. This interesting effect means that fermions may be created in significant numbers, even when the energy scale of the Universe has dropped below the natural energy scale of the fermions, leading to a recrudescence of their production.

The first studies of fermions coupled to a scalar condensate used a mode function approach, taking the condensate to be homogeneous. This was later extended, using techniques such as those pioneered in \cite{Borsanyi:2008eu}, to include inhomogeneities, as well as back-reaction. This led to a better understanding of how the homogeneous-condensate assumption underestimates the fermion production rate \cite{Berges:2010zv}. Indeed, the decay of oscillons was briefly described in \cite{Borsanyi:2008eu}, but from a different perspective that we pursue here.

The aim of this paper is to study the localized regions of oscillating condensate known as  oscillons \cite{Gleiser:1993pt},
and focus on how they emit massive fermions. If we view them as small lumps of preheating-phase physics, we might expect them to be able to emit heavy fermions rather efficiently, just as homogeneous preheating does. In this initial study we shall ignore the effects of back-reaction from the fermions onto the oscillon, and use techniques introduced by Cohen, Coleman, Georgi and Manohar \cite{Cohen:1986ct}, and followed up by \cite{Multamaki:1999an}.
In these the authors examined fermion emission from Q-balls \cite{Coleman:1985ki}, and calculated the particle production using a Bogoliubov-type analysis. In practical terms, the difference that makes the calculation more involved for oscillons, is that their amplitude is not constant, but this is also what leads to the fermion emission not being just a surface effect, unlike for Q-balls \cite{Cohen:1986ct}.

We study two models for a spherical, localised, oscillating condensate; one where the condensate has a broad almost homogeneous core, and one where the condensate profile is Gaussian. In our calculation the condensate is considered as an external source, and the dynamics of the oscillon itself are not examined, rather we look to see how fermions react to the condensate. Nevertheless, it is of interest to see what models produce such condensates, and these are described in the appendices. 

The layout of the paper is as follows: In section \ref{sec:model} we set out the basic spherical fermion ansatz, and describe the equations of motion, with the relevant mode functions for quantising the fermi field calculated in section \ref{sec:freeEvolution}. In section \ref{sec:includingOscillon} we describe how to include the oscillon, with our results presented in \ref{sec:results_flat} and \ref{sec:resultsGaussian} for the flat-topped and Gaussian oscillons respectively, with these oscillon models being described in appendices \ref{sec:flatTop} and \ref{sec:gaussian}. We conclude in section \ref{sec:conclusion}.

\section{Equations of motion}
\label{sec:model}

In this we will not assume that the effective fermion mass, $\mu$, is a constant, as the Yukawa coupling between the scalar and fermion will contribute to $\mu$. We shall, however, take it to be spherically symmetric, as the oscillon that contributes to $\mu$ will be taken to be spherical. This allows us, using standard spherical harmonic spinors $\Omega_{j,l_\pm,M}$ 
\cite{greiner:1998}, to write down an ansatz for the solution to the Dirac equation in terms of functions that depend solely on $t$ and $r$,
\ba\label{eq:spinorDecomposition}
\Psi_{j,M}(t,\underline r) &=& 
\left(
\begin{array}{c}
f_1(t,r)\Omega_{j,l_+,M}(\theta,\phi) \\
g_1(t,r) \Omega_{j,l_-,M}(\theta,\phi)
\end{array}
\right)
+
\left(
\begin{array}{c}
f_2(t,r)\Omega_{j,l_-,M}(\theta,\phi) \\
g_2(t,r) \Omega_{j,l_+,M}(\theta,\phi)
\end{array}
\right).
\ea
This is substituted into the Dirac equation to find the equations of motion for the functions $f_i(t,r)$ and $g_i(t,r)$, for example
\ba\label{eqs:fg_eqns1}
\dot f_1&=&i\mu f_1+g_1'-\frac{j-1/2}{r}g_1,\\\label{eqs:fg_eqns2}
\dot g_1&=&-i\mu g_1+f_1'+\frac{j+3/2}{r}f_1.
\ea

\section{Free fermion evolution}
\label{sec:freeEvolution}

In order to define a vacuum state for the fermion we need to know the mode functions for the Dirac equation in the absence of an oscillon. For this we consider constant-$\mu$, definite energy/wavenumber solutions of the Dirac equation, $\Psi^{\pm(\alpha)}_{j,M}(t,\underline r;k)$,
\ba\nonumber
\Psi^{+(\alpha)}_{j,M}=e^{-i\omega t} U_{j,M}^{(\alpha)}(\underline r;k),\quad
\Psi^{-(\alpha)}_{j,M}=e^{i\omega t} V_{j,M}^{(\alpha)}(\underline r;k),
\ea
where $U^{(\alpha)}$ and $V^{(\alpha)}$ are found to be
\ba\label{eq:U}
U^{(1,2)}_{j,M}(\underline r;k)&=&\frac{k}{\sqrt{\pi\omega}}
\left(
\begin{array}{c}
\sqrt{\omega-\mu}\;j_{l_\pm}(kr)\;\Omega_{j,l_\pm,M}\\
\pm i\sqrt{\omega+\mu}\;j_{l_\mp}(kr)\;\Omega_{j,l_\mp,M}
\end{array}
\right),\\\label{eq:V}
V^{(1,2)}_{j,M}(\underline r;k)&=&\frac{k}{\sqrt{\pi\omega}}
\left(
\begin{array}{c}
\pm i\sqrt{\omega+\mu}\;j_{l_\pm}(kr)\;\Omega_{j,l_\pm,M}\\
\sqrt{\omega-\mu}\;j_{l_\mp}(kr)\;\Omega_{j,l_\mp,M}
\end{array}
\right),\\\nonumber
\omega&=&+\sqrt{\mu^2+k^2}.
\ea
These play the role of the usual $U$ and $V$ plane-wave modes in Minkowski space Cartesian co-ordinates, with the $j_l(kr)$ being spherical Bessel functions. The particular factors appearing (\ref{eq:U}-\ref{eq:V}) are chosen such that 
\ba\nonumber
 \int d^3x\; U^{(\alpha)\dagger}_{j,M}(\underline r;k)U^{(\alpha)}_{j',M'}(\underline r;k')= 
\delta_{jj'}\delta_{MM'}\delta(k-k'),\\\nonumber
\ea
with a similar relation for the $V$, as well as the $U$ being orthogonal to the $V$ under the same inner product.

With the basic wave solutions found and normalized, we may consider quantization, and so we expand a general wave operator as
\ba\nonumber
\hat\Psi(t,\underline r) &=& \sum_{\alpha,j,M}\int dk\left\{ \hat b_{(\alpha)}(k,j,M)e^{-i\omega t}U^{(\alpha)}_{j,M}(\underline r;k) + \hat d^\dagger_{(\alpha)}(k,j,M)e^{i\omega t}V^{(\alpha)}_{j,M}(\underline r;k) \right \},
\ea
and find that taking the canonical anti-commutation relation for $\hat\Psi$ and its conjugate leads to
\ba\nonumber
\left\{\hat b_{(\alpha)}(k,j,M),\hat b^\dagger_{(\alpha')}(k',j',M')\right\}&=&\delta_{\alpha\alpha'}\delta_{jj'}\delta_{MM'}\delta(k-k'),
\ea
with a similar relation for $\hat d$. While it certainly proves useful to calculate the above mode functions, they do not constitute the basis of interest for our problem, rather we look to a basis that utilises incoming and outgoing modes. This we do in the next section.

\section{Including the oscillon}
\label{sec:includingOscillon}

\subsection{Fermion sector}
\label{sec:fermionSector}
Before we get to the solution in the presence of the oscillon we note from (\ref{eqs:fg_eqns1}-\ref{eqs:fg_eqns2}) and (\ref{eq:spinorDecomposition}) that time evolution couples $f_1$ to $g_1$, and it couples $f_2$ to $g_2$. In terms of the $U^{(\alpha)}$ and $V^{(\alpha)}$ modes this means $U^{(1)}$ is paired with $V^{(1)}$, while $U^{(2)}$ is paired with $V^{(2)}$. So, we may introduce another basis of functions (the scattering basis) that are also solutions for constant $\mu$. One of these basis functions is, schematically,
\ba\label{eq:chi1}
\chi_1&\sim&e^{-i\omega t}U^{(1)}(\underline r,k | h^{(2)})+e^{-i\omega t}R_1U^{(1)}(\underline r,k | h^{(1)})+e^{i\omega t}T_1V^{(1)}(\underline r,k | h^{(1)}).
\ea
This solution has an extra argument for $U^{(1)}$ and $V^{(1)}$, namely $h^{(1,2)}$, which is to indicate that the $j_l(kr)$ of (\ref{eq:U}-\ref{eq:V}) are to be replaced by spherical Bessel functions of the third kind ($h^{(1,2)}_n=j_n\pm iy_n$). This means that they are still solutions to the equations of motion, but are only regular away from the origin. In fact, we will only be interested in this form at radii $r\gg R_{osc}$, which is where we shall make our measurements of particle number. Note that at large radius, $h^{(1)}\sim \frac{1}{kr}e^{ikr}$ and so corresponds to an outgoing wave, while $h^{(2)}$ corresponds to an ingoing wave. We now think about wavepackets formed from such solutions, rather than the pure frequency modes, and then we see that at early times (\ref{eq:chi1}) should be thought of as an ingoing $U^{(1)}$ mode, and at late times it is a combination of outgoing $U^{(1)}$ and outgoing $V^{(1)}$, in proportion determined by $R_1$ and $T_1$. 

Unlike the Q-balls studied in \cite{Cohen:1986ct}, oscillons have varying amplitude, which makes the situation more involved, as the time dependence cannot be solved by a simple phase dependence. In practise, this means that even if the ingoing wave has a single frequency, the outgoing wave does not, and so contains a spectrum of wavenumbers. The scattering basis for large radius, $r\gg R_{osc}$, is then
\ba\label{eq:chi2}
\chi_{(\alpha)}(t,\underline r; k_{in},j,M)&=&e^{-i\omega_{in} t} U^{{(\alpha)}}_{j,M}(\underline r;k_{in}\;|\;h^{(2)}) \\\nonumber
	&~&+\int dk\; e^{-i\omega t}R^{(\alpha)}(k,k_{in})U^{{(\alpha)}}_{j,M}(\underline r;k\;|\;h^{(1)})+\int dk\; e^{i\omega t}T^{(\alpha)}(k,k_{in})V^{{(\alpha)}}_{j,M}(\underline r;k\;|\;h^{(1)}),\\
	%
	%
\label{eq:zeta2}
\zeta_{(\alpha)}(t,\underline r;k_{in},j,M)&=&e^{i\omega_{in} t} V^{{(\alpha)}}_{j,M}(\underline r;k_{in}\;|\;h^{(2)}) \\\nonumber
	&~&+\int dk\; e^{i\omega t}\tilde R^{(\alpha)}(k,k_{in})V^{{(\alpha)}}_{j,M}(\underline r;k\;|\;h^{(1)})+\int dk\; e^{-i\omega t}\tilde T^{(\alpha)}(k,k_{in})U^{{(\alpha)}}_{j,M}(\underline r;k\;|\;h^{(1)}),
	%
\ea
and so we expand a general wave operator as,
\ba\nonumber
\Psi(t,\underline r) = 
&&\sum_{\alpha,j,M}\int dk_{in}\left\{ \;\hat b_{in(\alpha)}(k_{in},j,M)\chi_{(\alpha)}(t,\underline r;k_{in},j,M) + \hat d^\dagger_{in(\alpha)}(k_{in},j,M)\zeta_{(\alpha)}(t,\underline r;k_{in},j,M)  \right \}.\\\label{eq:Psi_chi_zeta}
\ea
At this point it is useful to understand why the operators have been labelled with the subscript ``$in$". Thinking about the wave operator in terms of wave packets, we have that in the far past only the incoming part of the scattering basis survives (the terms depending on $h^{(2)}$) and so the wave operator is indeed a sum of ingoing waves, and the operators $(d_{in}^\dagger)$, $b_{in}^\dagger$ acquire the interpretation as creation operators for ingoing (anti-)particles. 

At late times, the wave operator will be composed solely of outgoing particles, so at large radius we would write
\ba\nonumber
\Psi(t,\underline r) &=& \sum_{\alpha,j,M}\int dk\;\left\{ \;\hat b_{out(\alpha)}(k,j,M)e^{-i\omega t}U^{(\alpha)}_{j,M}(\underline r;k\;|\;h^{(1)}) + \hat d^\dagger_{out(\alpha)}(k,j,M)e^{i\omega t}V^{(\alpha)}_{j,M}(\underline r;k\;|\;h^{(1)})  \right \},\\\label{eq:psiOut}
\ea
and the operators $(d_{out}^\dagger)$, $b_{out}^\dagger$ are creation operators for outgoing (anti-)particles. However, we know what the late time, large radius form is from  (\ref{eq:Psi_chi_zeta}) and (\ref{eq:chi2}-\ref{eq:zeta2}), so we compare these (in the late time, large radius limit) with (\ref{eq:psiOut}) to find
\ba\nonumber
\hat b_{out(\alpha)}(k,j,M) &=&\int dk_{in}\left\{R_{(\alpha)}(k,k_{in})\hat b_{in(\alpha)}(k_{in},j,M) + \tilde T_{(\alpha)}(k,k_{in})\hat d^\dagger_{in(\alpha)}(k_{in},j,M)\right\},
\ea
and a similar relation holds for $\hat d_{in}$ and $\hat d_{out}$ . These give the Bogoliubov transformation between asymptotic in/out creation and annihilation operators. 

The number of fermions emitted by the oscillon are found by starting the system in the vacuum state $|0\rangle$, defined by having no incoming fermions or anti-fermions,
\ba
b_{in}|0\rangle &=& 0,\qquad d_{in}|0\rangle = 0,
\ea
and we count the number of fermions per unit $k$-space at wavenumber $k$, at late times, by evaluating
\ba\nonumber
N_{out}(k)&=&\sum_{\alpha,j,M}\langle 0|b^\dagger_{out(\alpha)}(k,j,M)b_{out(\alpha)}(k,j,M)|0\rangle,\\\label{eq:N_k}
	&=&\sum_{\alpha,j}(2j+1)\int dk_{in}|\tilde T_{(\alpha)}(k,k_{in})|^2.
\ea
This term diverges, but the divergence is to be expected as the oscillon is evolving without back-reaction and continues producing particles forever.
In practise one is really interested in $\frac{\Delta N_{out}}{\Delta\tau}$, where $\Delta N_{out}$ is the number of fermions produced in time interval $\Delta\tau$. The effort that goes into finding the rate of (anti-)particle production, therefore, amounts to determining $(T^{(\alpha)})$, $\tilde T^{(\alpha)}$. This is achieved by noting that in terms of the functions $f$ and $g$ appearing in (\ref{eqs:fg_eqns1}-\ref{eqs:fg_eqns2}), we may use (\ref{eq:chi2}-\ref{eq:zeta2}) along with (\ref{eq:U}-\ref{eq:V}) to find the following large radius behaviour for $f_1(t,r)$
\ba
f_1(t,r;k_{in}) &=& \frac{k_{in}\sqrt{\omega_{in}-\mu}}{\sqrt{\pi\omega_{in}}}e^{-i\omega_{in}t}h^{(2)}_{l_+}(k_{in}r)\\\nonumber
	&~&+\int dk\frac{k\sqrt{\omega-\mu}}{\sqrt{\pi\omega}}R^{(1)}(k,k_{in})e^{-i\omega t}h^{(1)}_{l_+}(kr)\\\nonumber
	&~&\int dk\frac{ik\sqrt{\omega+\mu}}{\sqrt{\pi\omega}}T^{(1)}(k,k_{in})e^{i\omega t}h^{(1)}_{l_+}(kr),
\ea
with analogous results for $f_2$, $g_1$ and $g_2$. We see, therefore, that if we observe $f_1$ at a given radius $r=\rho\gg R_{osc}$, over some time $\Delta\tau$, then we may perform a temporal Fourier transform $\int dt\; e^{-i\Omega t}f_1(t,r=\rho;k_{in})$
to find $T^{(1)}(K=\sqrt{\Omega^2-\mu^2},k_{in})$. Having set up the formalism, we now need a condensate to create the fermions, and this we address in the next section.

\section{Results: flat-topped oscillons}
\subsection{flat-topped oscillons}
\label{sec:results_flat}
The scalar sector for the flat-topped oscillons is set out in App. \ref{sec:flatTop}, and we couple this to a fermion which has free mass $m_{0}$, via a Yukawa term ${\cal L}_{Yuk}=\xi\phi\bar\psi\psi$, giving an effective fermion mass of 
\ba
\mu&=&m_0+\xi\phi.
\ea
One of the important observations made in \cite{Giudice:1999fb}\cite{Greene:1999hw} is that the production of massive fermions due to an oscillating scalar condensate can be efficient if the effective mass of the fermions goes through zero at some point during the condensate cycle; this requires $\phi_{max} \gtrsim m_0/\xi$. Moreover, the coherent nature of the condensate allows for fermions heavier than the scalar to be produced \cite{Giudice:1999fb}\cite{Greene:1999hw}.

The parameters we chose were as follows: $\xi=1.0$, $m=0.5m_0$, $\lambda=0.1$, $g=4\lambda^2$, $\Phi_0^2=\Phi_c^2-10^{-4}$, which leads to an oscillon of radius $\sim 30m_0^{-1}$ and energy $\sim30,200m_0$, with period of oscillation $T_{osc}\sim13.8m_0^{-1}$. For the simulation of (\ref{eqs:fg_eqns1}-\ref{eqs:fg_eqns2}), we  used a lattice of $40,000$ points, with a spatial step of $dr=0.02m_0$, and a temporal step size of $dt=dr/5.0$. We should note that the mass of the scalar that forms the condensate is lower than the free mass of the fermion.

\begin{figure}
  	\centering
    	\includegraphics[width=0.8\textwidth]{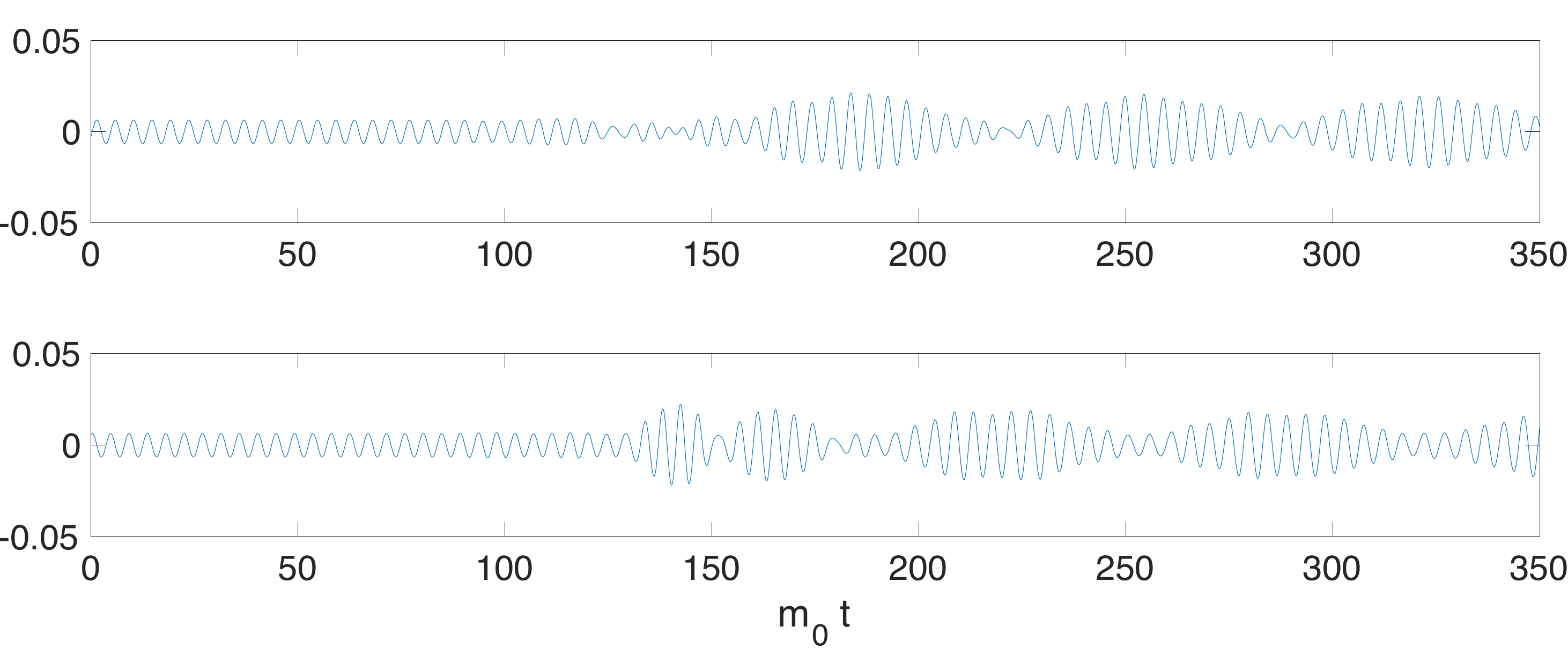}
  	\caption{An example of $f_1(t,r=\rho)$ for $j=4\frac{1}{2}$, $\rho=100m_0^{-1}$ and $k_{in}=m_0$ for the flat-topped oscillon. The top figure is the real part of $f_1$ and the lower is the imaginary part.}\label{fig:f_rho_jeq4p5_keq1}
\end{figure}

In order to evaluate the emission rate we need the Fourier transform of $f_1$ evaluated at a fixed radius, $\rho$. We give an example of the mode function $f_1(t,r=100m_0^{-1})$ in Fig. \ref{fig:f_rho_jeq4p5_keq1}. Here we see that not much happens until $m_0t\sim120$, at which point there is a burst of activity, before settling down to a periodic motion. The burst  around $m_0t\sim 120\to 150$ is due to the osciillon effectively being suddenly switched on  at the start of the simulation. We let this burst pass before we start taking the Fourier transform, for which we use $180<m_0t<680$.

\begin{figure}
  	\centering
    	\includegraphics[width=0.8\textwidth]{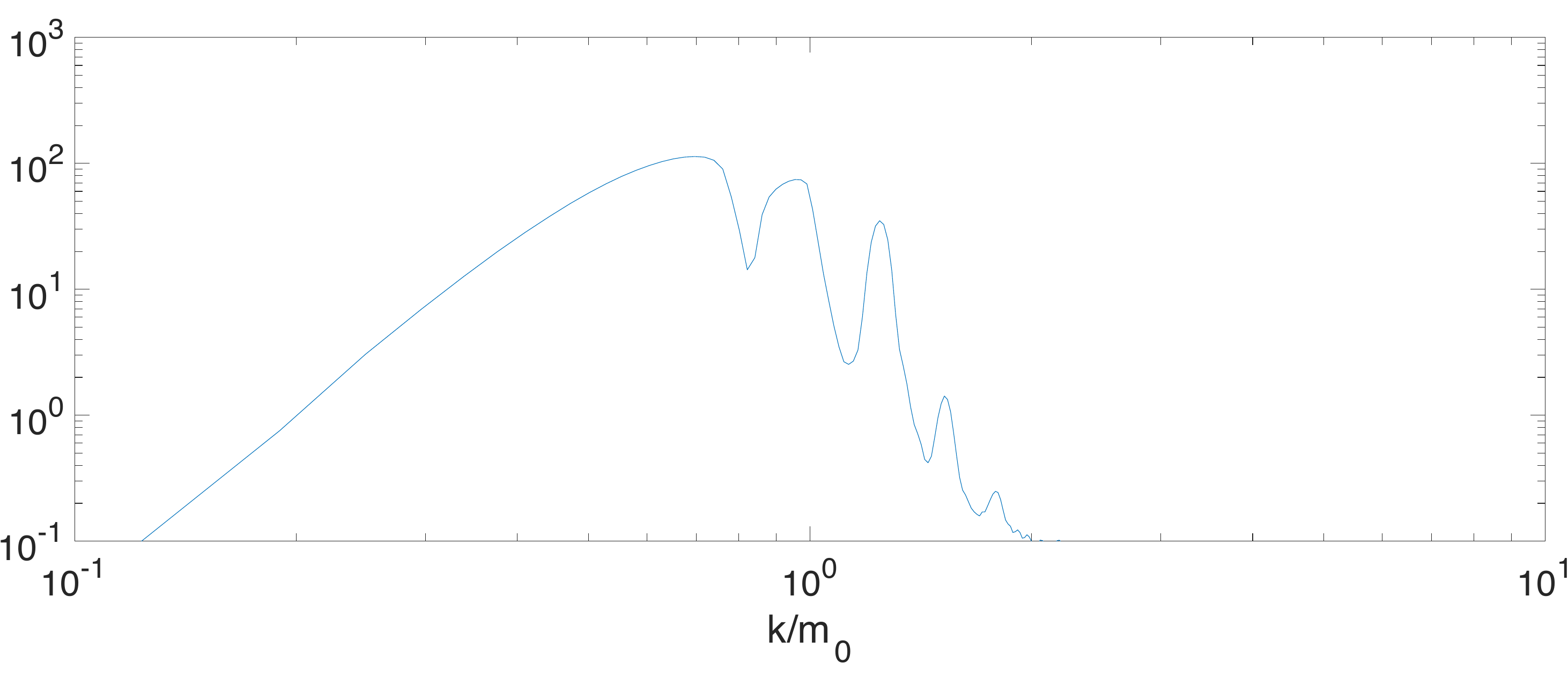}
  	\caption{A plot showing $\frac{\Delta N(k)}{\Delta \tau}$ for the flat-topped oscillon example in the text.}\label{fig:dNdk}
\end{figure}

The formalism laid out above now allows us to evaluate the number of fermions per $k$-space interval emitted per unit time, $\frac{\Delta N(k)}{\Delta \tau}$, and this is shown in Fig. \ref{fig:dNdk} - for this example we found that the sum in (\ref{eq:N_k}) had converged by $j_{max}=29\frac{1}{2}$, which is what we use for the figure. The plot indicates a number of resonance peaks, consistent with the picture one has for a homogeneous condensate \cite{Giudice:1999fb}.
\begin{figure}
  	\centering
    	\includegraphics[width=0.8\textwidth]{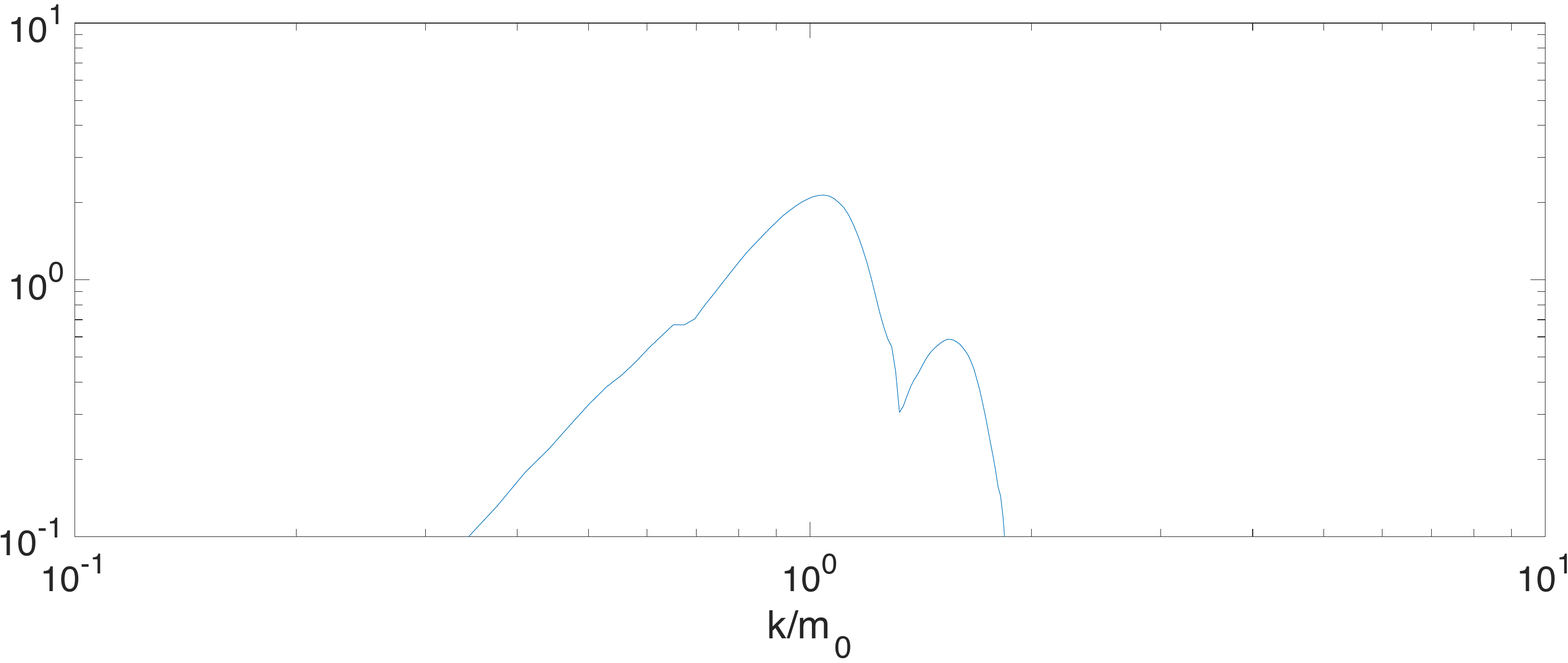}
  	\caption{A plot showing $\frac{\Delta N(k)}{\Delta \tau}$ for the Gaussian oscillon example in the text.}\label{fig:dNdkGaussian}
\end{figure}
As well as the spectrum of emitted particles it is of interest to know how many fermions are emitted in total over one cycle, which may be found by performing an integral over $k$-space of the spectrum,
\ba
\frac{\Delta N}{\Delta \tau} &=&\int dk\;\frac{\Delta N(k)}{\Delta\tau},
\ea
and in our example yields $\frac{\Delta N}{\Delta \tau}\simeq 50 m_0$. So, over a single cycle of duration $T_{osc}$ our flat-topped oscillon emits $\sim 690$ fermions, constituting about 2\% of its energy budget per  cycle.

\begin{figure}
  	\centering
    	\includegraphics[width=0.8\textwidth]{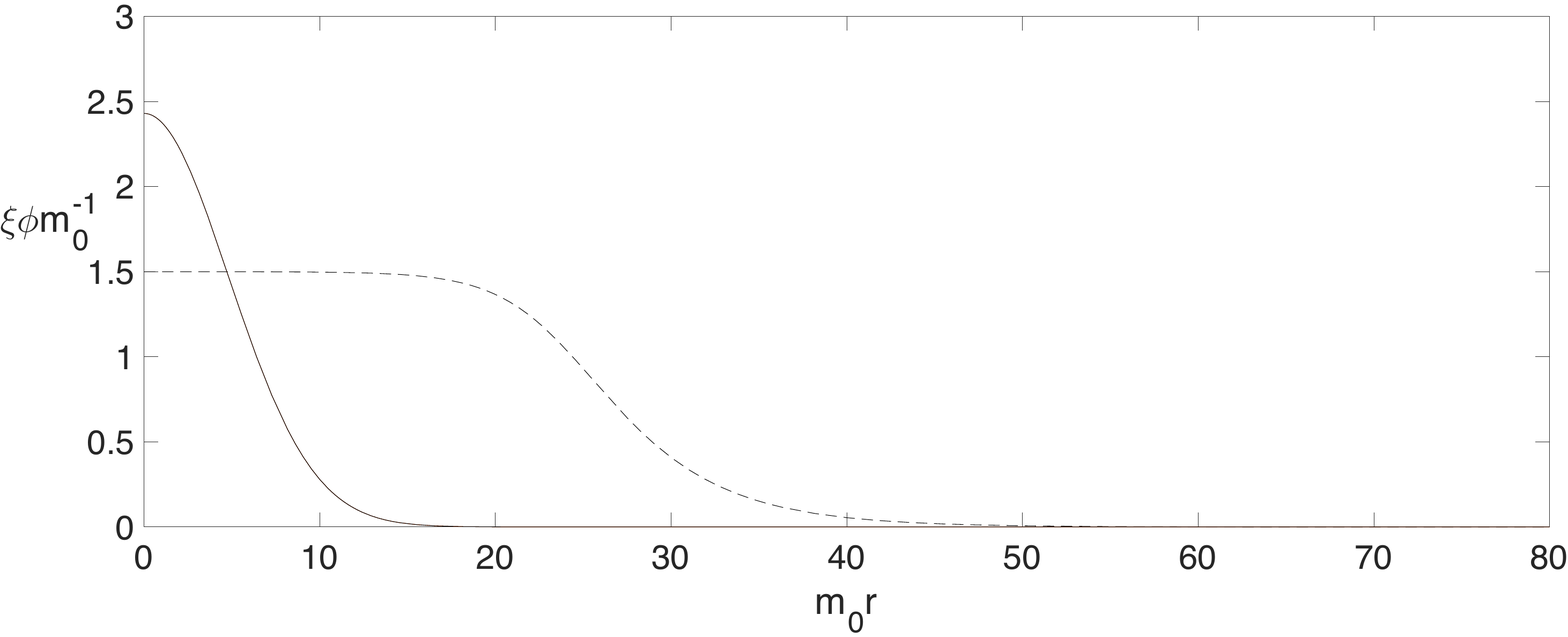}
  	\caption{A plot showing the scalar field profiles for the flat-topped oscillon (dashed line) and the Gaussian oscillon (solid line) examples in the text. Note that they are presented with the Yukawa factor included.}\label{fig:profiles}
\end{figure}

\subsection{Gaussian oscillons}
\label{sec:resultsGaussian}
The more commonly studied class of oscillons, laid out in App. \ref{sec:gaussian}, are those with an approximately Gaussian profile. For our simulations in this case, we take a Yukawa coupling of $\xi=0.1$, with again $m=0.5m_0$. The parameters given in App. \ref{sec:gaussian} lead to an oscillon with period $T_{osc}\simeq 14.3m_0^{-1}$, core amplitude of $A_B=24.3m_0$, and radius of $R=6.8m_0^{-1}$. Numerical simulations using the same lattice parameters as for the flat-topped oscillon yields the production rate spectrum shown in Fig. \ref{fig:dNdkGaussian}, and the total production rate $\frac{\Delta N}{\Delta \tau}\simeq 1.3 m_0$. So, over a single cycle of duration $T_{osc}$ our oscillon emits $\sim 19$ heavy fermions, constituting about 0.06\% of its energy budget. Such a reduction is to be expected given that for our parameters the Gaussian oscillon is smaller than the flat-topped oscillon, as may be seem from Fig. \ref{fig:profiles}. Even though this is a small fraction of the oscillon's energy, one should bear in mind that oscillons can live for many thousands of oscillations.
\section{Conclusion and discussion}
\label{sec:conclusion}
Based on the observation from preheating studies that a homogeneous oscillating scalar condensate is able to produce heavy fermions, and do so more efficiently than bosons \cite{Giudice:1999fb}, we have initiated a study into the fermion production due to mini preheating regions, oscillons. The basic formalism for calculating the fermion emission was laid out, and an example given using an approximate analytic form for the oscillon solution. It was shown, by example, that despite the scalar field being lighter than the free-fermion mass, the coherent nature of the oscillon allowed fermions to be produced, and in significant numbers, given that oscillons can live for thousands of oscillations. The possibility that such oscillons could form at "low" energies and yet produce heavy fermions raises interesting questions for baryogenesis, where such heavy fermions may, for example, be heavy right-handed Majorana neutrinos. These would break the $B-L$ symmetry of the standard model, and their decay causes a lepton asymmetry, which leads to a baryon asymmetry due to sphaleron processes. It is also amusing to consider the possibility that we may one day be able to manipulate a Higgs condensate into performing localised oscillations, thereby producing particles heavier than the scalar itself.   

There are still a number of things to be done, one of which is to acquire a more complete picture of how the fermion production is affected by the oscillon's properties, such as its amplitude, size or frequency. For example, if $\phi_{max}<m_0/\xi$ then fermion production may cease, as their effective mass would never vanish. Another important aspect is back-reaction, whereby the fermion dynamics alters the behaviour of the oscillon. This has proved an important effect in cosmological preheating, where is has been noted that neglecting back-reaction underestimates the number of fermions produced \cite{Berges:2010zv}. The fermion coupling may also change the way that oscillons form, with possible incipient fermion production happening during the formation process. This, however, will require large-scale simulations to gain a full understanding, and is something left for future studies.
 
\vspace{0.4cm}

\noindent {\bf Acknowledgments:} We would like to thank STFC for financial support under grant ST/L000393/1.

\appendix

\section{Scalar sector: flat-topped oscillons}
\label{sec:flatTop}
Much has been written about oscillons since their discovery \cite{Bogolyubsky:1976nx} and rediscovery \cite{Gleiser:1993pt}. They are localized lumps of oscillating scalar-field condensate, made quasi-stable by non-linearities in the field equations, and may appear, for example, at phase transitions \cite{Kolb:1993hw}.
Due them not being perfectly stable, they radiate energy \cite{Segur:1987mg}
and have a lifetime with a significant dependence on the spatial dimension \cite{Saffin:2006yk}.
Ultimately, one is interested in treating the fields quantum mechanically, and results in that direction may be found in \cite{Hertzberg:2010yz}.
Their relevance to post-inflationary dynamics is examined in \cite{Amin:2011hj}, and it is found that in a large class of models that exhibit preheating, oscillons can come to dominate the energy density. In this paper we will not concern ourselves too much with the scalar sector, and will not include the back-reaction of the fermions onto the oscillon, leaving this for a future study. We shall follow \cite{Amin:2010jq} and take an analytic profile for the oscillons of the form
\ba\nonumber
\phi(t,r) &=& \Phi_0\sqrt{\frac{\lambda}{g}}\sqrt{\frac{1+u}{1+u\cosh(2\alpha\lambda x/\sqrt{g})}}\;\cos(\omega mt),
\ea
which corresponds to a localized, oscillating field distribution.
This profile is derived as an oscillon for the following Lagrangian density,
\ba\label{eqn:scalarLagrangian}
{\cal L}_{scalar,\;flat} &=& -\frac{1}{2}\del_\mu\phi\del^\mu\phi-\frac{1}{2}m^2\phi^2+\frac{\lambda}{4}\phi^4-\frac{g}{6}\phi^6,
\ea
with
\ba\nonumber
\alpha^2&=&\frac{3}{8}\Phi^2_0-\frac{5}{24}\Phi^4_0,\;\alpha_c=\sqrt{27/160},\omega^2=1-\alpha^2\frac{\lambda^2}{m^2g},\\\nonumber
u&=&\sqrt{1-(\alpha/\alpha_c)^2},\Phi_0=\Phi_c\sqrt{1-u},\;\Phi_c=\sqrt{9/10}.
\ea
Strictly speaking, such a solution is valid only in one spatial dimension in the limit \mbox{$g\gg\lambda/m^2$}.
The precise results of fermion emission will, of course, be affected by the detailed structure of the oscillon, but we are interested in how fermions react to a generic localized oscillating lump of scalar condensate, and will leave more detailed studies for future work.

Unfortunately, the region of parameter space required for flat-topped oscillons in this model is not radiatively stable. This is seen by requiring the effective eight-point vertex, brought about by the one-loop diagram with two $\phi^6$ vertices, to be smaller than the tree-level diagram with eight external lines (a $\phi^4$ vertex connecting a $\phi^6$ vertex via a propagator), and leads to a cut-off of $\Lambda=\sqrt{\lambda/g}$. For the analytic flat-topped solution to be valid we required $g\gg\lambda/m^2$ or, equivalently, $m\gg \sqrt{\lambda/g}$, which means the scalar mass has to be above the cut-off.

However, we are only considering the behaviour of fermions in the background of such an extended object, and have not considered the scalar field as dynamical. As such, the Lagrangian that leads to the profile has no bearing on the calculation, and we are simply taking this profile as a prototype for a localized oscillating condensate, and  simply consider it as an external source.

\section{Scalar sector: Gaussian oscillons}
\label{sec:gaussian}
The more common class of oscillons are those with Gaussian-like profiles, and for this we may consider the following Lagrangian
\ba\label{eqn:scalarLagrangianGaussian}
{\cal L}_{scalar,\;Gaussian} &=& -\frac{1}{2}\del_\mu\phi\del^\mu\phi-\frac{\lambda}{4}\left(\left[\phi-\eta\right]^2-\eta^2\right)^2,
\ea
which is simply the familiar Mexican-hat shaped potential, shifted so that the vacua are located at $\phi_{vac}=(0,2\eta)$. This model has oscillons of the approximate form \cite{Gleiser:2008ty}
\ba
\phi(t,r)&=&A_B\exp\left( -r^2/R^2 \right)\cos(\omega_Bt),\\
R&\simeq& \frac{2.42}{\sqrt\lambda\eta},\quad \omega_B\simeq 1.25\sqrt{\lambda}\eta,\quad A_B\simeq 1.54\eta,
\ea
and the energy of such a condensate is
\ba
E_{Gaussian}\simeq 41.3\frac{\eta}{\sqrt\lambda}.
\ea
The mass of the scalar in this model is $m=\sqrt{2\lambda}\eta$, which we fix to be half the fermion vacuum mass, $m=\frac{1}{2}m_0$, giving $E_{Gaussian}\simeq\frac{41.3}{2\sqrt{2}\lambda}m_0$. The coupling constant $\lambda$ is fixed by requiring the Gaussian oscillon to have a comparable energy to the flat-topped oscillon ($E_{flat-topped}\simeq30,200m_0$), and we take $\lambda=5\times 10^{-4}$, which is safely within the perturbative regime and leads to $E_{Gaussian}\simeq 29,200m_0$.

\end{document}